# Compressive Fourier Transform Spectroscopy

**Ori Katz, Jonathan M. Levitt, and Yaron Silberberg**

*Department of Physics of Complex Systems, Weizmann Institute of Science, Rehovot, 76100 Israel*

*ori.katz@weizmann.ac.il*

**Abstract:** We describe an approach based on compressive-sampling which allows for a considerable reduction in the acquisition time in Fourier-transform spectroscopy. In this approach, an N-point Fourier spectrum is resolved from much less than N time-domain measurements using a compressive-sensing reconstruction algorithm. We demonstrate the technique by resolving sparse vibrational spectra using <25% of the Nyquist rate samples in single-pulse CARS experiments. The method requires no modifications to the experimental setup and can be directly applied to any Fourier-transform spectroscopy measurement, in particular multidimensional spectroscopy.

## 1. Introduction

In Fourier transform (FT) spectroscopy an N-point spectrum is resolved by Fourier transforming a set of equally-spaced *N* time-domain measurements. FT spectroscopy benefits from improved SNR compared to direct spectral measurements (the Jacquinot-Felget advantage) and has been demonstrated useful in a variety of spectroscopic applications ranging from infrared spectroscopy (FT-IR) to nuclear magnetic resonance, including spontaneous and coherent Raman spectroscopy, mass spectrometry and electron spin resonance spectroscopy. The underlying principle behind FT spectroscopy is the familiar Shannon-Nyquist (also Whittaker-Kotel´nikov) sampling theorem, which states that in order to reconstruct a signal which is band-limited to a bandwidth B (in Hz), it is sufficient to sample the signal at 2B samples per second (the Nyquist rate). Sampling a signal for a finite time T, will result in a total of N=2BT samples. The discrete Fourier transform of such sampled signals will then contain N frequency coefficients, ranging from $f_{min}=-B$ to $f_{max}=B$ with a spectral resolution of $\Delta f = 1/T = 2B/N$.

Relying only on the Nyquist/Shannon theorem one may be led to believe that exactly N time-domain samples are needed to reconstruct an N-point bandwidth-limited spectrum. However, any prior information on the structure of the to-be-measured spectra can significantly reduce the number of measurements required for faithful spectral reconstruction. A careful inspection of the Raman vibrational spectrum of simple molecules reveals that such spectra are *sparse*, meaning that the number of nonzero spectral components, K, is very small relative to the number of total spectral bins, N, available in the measurement resolution. Although the total number of active frequencies in such signals is small, the material specific frequencies and their corresponding amplitudes are not known. Consequently, one cannot simply restrict the bandwidth a priori, and using only the Nyquist/Shannon theorem still requires that all N time-domain samples are to be measured.

The study of sparse signals within the framework of the emerging field of c*ompressive sampling* (also *compressed/compressive sensing*, CS [1,2]) shows that one can reconstruct an N-point signal which is composed of K<<N nonzero frequency components using only M<<N time-samples, yielding a *sub-Nyquist sampling criterion*: $M>2K\cdot\log(N/M)$ [3]. Moreover, these M time-domain samples do not have to be carefully chosen; almost any random sample set of this size will enable a faithful reconstruction. In addition, CS reconstruction is robust to measurement noise and is applicable to signals which are not exactly sparse but are *compressible*, i.e. the sorted magnitudes of the signal spectral coefficients decay quickly and do not have to vanish completely [1], features which are essential under practical experimental conditions.

In this work we demonstrate CS undersampling approach for resolving sparse vibrational spectra in Fourier-transform single-pulse Coherent Anti-Stokes Raman scattering (CARS) spectroscopy [4-5]. We experimentally resolve vibrational spectra of simple molecules using < 30% of the Nyquist limit time-domain measurements.

## 2. Numerical results

Numerically simulated CARS signals obtained by the impulsive selective excitation scheme presented in [4] were used to test the CS reconstruction algorithm. A time-domain measurements trace of a toluene sample composed of three vibrational lines (522cm$^{-1}$, 787cm$^{-1}$ and 1005cm$^{-1}$) was simulated numerically using the theory appearing in [5], assuming 3cm$^{-1}$ Raman linewidths. The time domain trace was composed of N=281 equally spaced measurements (temporal delays from 200fs to 3000fs in 10fs steps). From this N-point trace, M<N points were *randomly chosen* and were used as the input for the CS reconstruction algorithm described in [6]. The CS reconstruction algorithm works by looking for the signal having the sparsest spectrum which still fulfills the M time-domain measurements made. We used sparsity in the discrete cosine-transform basis, which has similar sparsity to the FT basis and is composed of real (rather than complex) coefficients. The results comparing the full N=281 point Fourier transform to the undersampled M=60 point Fourier transform and CS reconstruction are presented in Fig.1. We found that for resolving the sample Raman lines M>~60 samples were required (x5 times sub-Nyquist). A result which is in reasonable agreement with the CS sampling criterion for a K=10 sparse signal [1,3]. All Fourier transforms were made on the signal after windowing with a Hamming window and subtraction of a DC frequency component.

## 3. Experimental Results

We tested the CS undersampling approach on experimental data from single-pulse CARS experiments [4]. Time-domain measurements for chloroform and toluene were taken using the scheme described in [4]. The laser source was a homemade Ti:Sapphire oscillator with spectral width of 60nm FWHM. The pulses were shaped using a 4-f pulse shaper apparatus based on a Holoeye AG HEO1080P spatial-light modulator. The shaped pulses of 20mW average power were focused on the sample using a 0.45 NA objective. Figure 2 shows the experimental results for chlroform using the full N=281 data-point trace (200fs to 3000fs) and the CS reconstruction using only M=65 randomly chosen measurements. The 60 sampled points were randomly chosen from the temporal delays in the range < 2000fs as these had superior signal-to-noise than the larger delay values, due to vibrational motion decay.

Although a shorter time-trace was used, no reduction in spectral resolution was observed, indicating spectral ‘super-resolution’, analogous to the spatial super-resolution recently reported in [10]. Before the analysis, a linear fit was subtracted from the measured experimental data-points to account for diffraction losses in the pulse shaper.

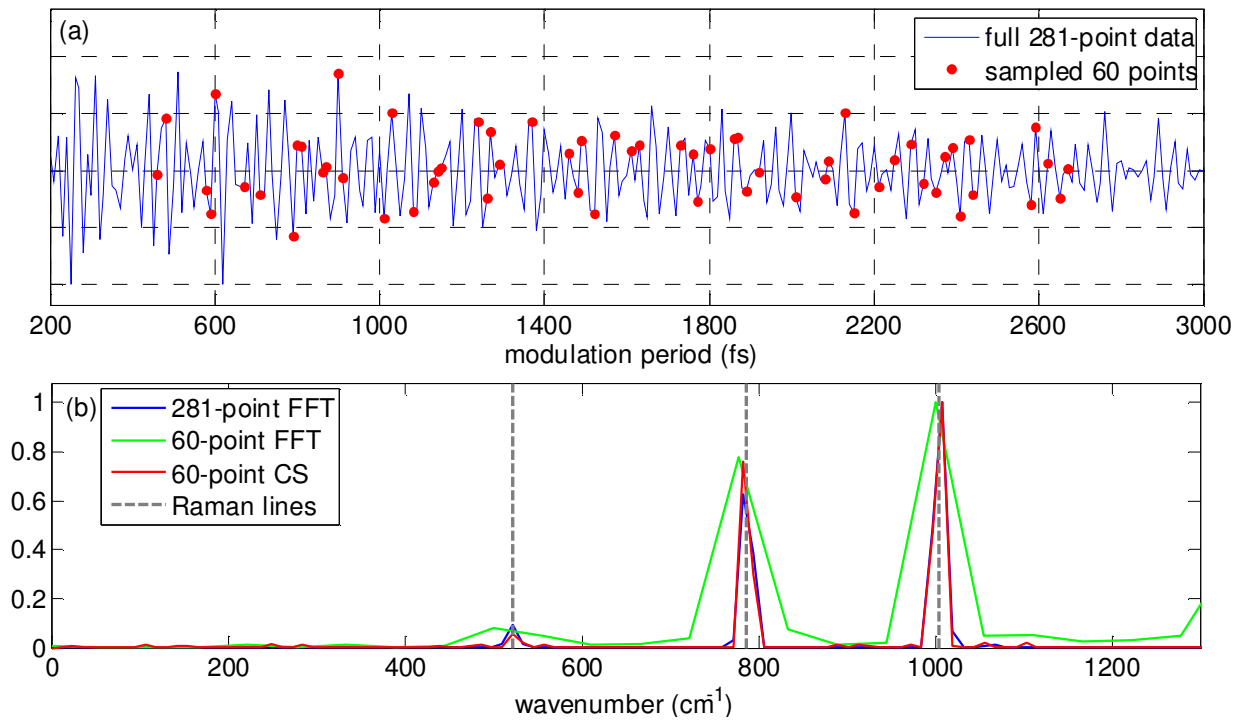

**Figure 1. Numerical results of compressive-sensing Fourier-transform vibrational spectroscopy using under-sampled time-domain data. (a) raw time-domain trace for toluene, containing 281 points (blue) and the randomly sampled 60 points (red). (b) resolved vibrational spectrum from full data (blue), 60-point data using standard FT (green) and 60-point data using CS reconstruction (20% the Nyquist limit, red).**

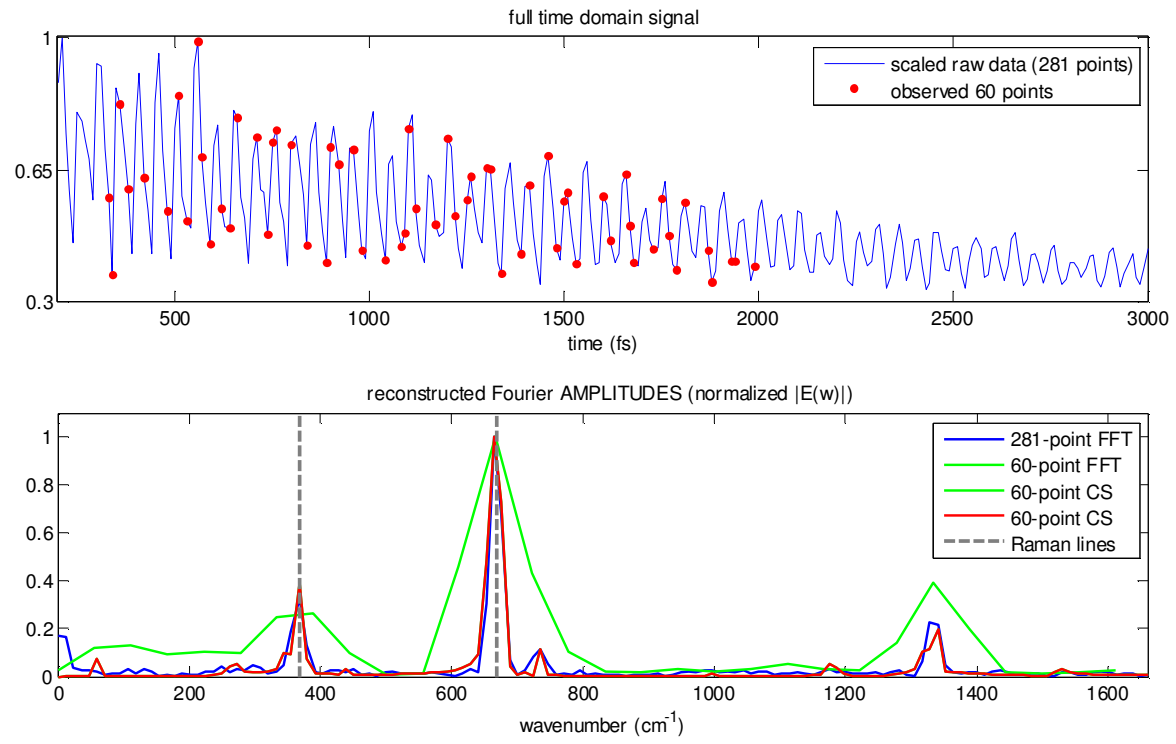

**Figure 2. Experimental results of compressive-sensing Fourier-transform vibrational spectroscopy using under-sampled time-domain data. (a) raw time-domain trace for chloroform, containing 281 points (blue) and the randomly sampled 60 points (red). (b) resolved vibrational spectrum from full data (blue), 60-point data using standard FT (green) and 60-point data using CS reconstruction (red).**

## 4. Discussion

We propose the use of a CS approach in Fourier-transform spectroscopy and experimentally demonstrate that CS allows a 4-folds reduction in the acquisition time of sparse vibrational spectra, an attractive feature in applications such as remote-sensing and vibrational microscopy. The technique can be directly applied to any Fourier-transform spectroscopy measurement, particularly multidimensional spectroscopy, where the acquisition times are long and the multidimensional spectra are usually sparse in cases where shifting to the rotating-frame is not used [7-9]. Furthermore, this CS recovery of two-dimensional data has been recently demonstrated in the context of optics for ghost-imaging [11]. For complex non-sparse spectra, the compressibility in other transform bases can be exploited and was demonstrated for spectro-polarimetric data using sparsity in wavelet decomposition [12] and for reflectance and scattering-polarization spectra using principle component analysis (PCA) decomposition [12,13]. In these works, the measurements were obtained by spectral multiplexing techniques similar in principle to Hadamard spectroscopy, and not through Fourier-transform time-sampling as presented in this work.